# Very slow surface plasmons: theory and practice

V. S. Zuev (1) and G. Ya. Zueva (2)

(1) The P. N. Lebedev Physical Institute of RAS
email: vizuev@sci.lebedev.ru
(2) The A. M. Prokhorov General Physics Institute of RAS

The paper is of a methodological character and has as a goal to give a brief description of the concept of theory and practical application of very slow optical plasmons. They exist on the metal-dielectric boundaries, namely, on very thin metal films and fibers and as standing waves on metal spheres and ellipsoids. The material presented in the paper features by widening the common concepts of electromagnetic modes of various spaces, of the probability of spontaneous emission, of creation of optical images, of the limits of optical focusing, and of the photon linear momentum. All mentioned studies are completed in the recent years. The problem of the photon momentum in a dielectric medium was the topic of irreconcilable disputes for 100 years starting in the time of appearing Minkowski and Abraham famous papers. Various practical applications are surveyed – the experiments with a great intensification of an atom spontaneous emission into a plasmonic field mode of a metal nanoparticle, the experiments on focusing of optical radiation into a spot that substantially smaller than a diffraction limited spot, a so called near perfect Pendry lens that produces the image with details that substantially smaller than defined by diffraction, and, lastly, the concept of hundredfold and more magnification of a photon mechanical linear momentum in a plasmon.

The work completed is supported by RFBR, grants Nos. 05-02-19647, 07-02-01328.

# Очень медленные поверхностные плазмоны: теория и практика

В.С.Зуев [1], Г.Я.Зуева [2]

[1] Физический институт им. П.Н.Лебедева РАН
[2] Институт общей физики им. А.М.Прохорова РАН

Статья носит методический характер и имеет целью дать представление о теории и практике применения очень медленных оптических плазмонов, существующих на границах раздела металл-диэлектрик - на тонких пленках металла и на тонких металлических нитях, а также на сферах и эллипсоидах, на которых возбуждаются стоячие волны-плазмоны. Представляемый материал отличается тем, что расширяет стандартные представления об электромагнитных модах пространств, о вероятностях спонтанного излучения, о построении оптических изображений, об оптической фокусировке, об импульсе фотона. Все эти исследования являются достижениями последних лет. Вопрос об импульсе фотона в среде являлся предметом непримиримых споров на протяжении почти ста лет, начиная со знаменитых Минковского и Абрагама. Рассмотрены практические применения – опыты по значительной интенсификации спонтанного излучения атома в плазмонную моду поля наночастицы, опыты по фокусировке оптического излучения в пятно размером существенно меньше дифракционного, так называемая почти идеальная линза Пендри, которая строит изображение с деталями, которые существенно меньше длины волны и, наконец, сильно увеличенный (десяти-стократно и более) импульс фотона в плазмоне.

Предлагаемый ниже материал в заметной степени основан на идеях Пендри /1/, Яблоновича /2/ и Поля /3/. Также используются основные результаты из /4,5,6/ и из наших работ /7-10/. Данные о проницаемости металлов содержатся в /11/.

Понятие поверхностного плазмона возникает как решение волнового уравнения для электромагнитного поля. Хорошо известным решением волнового уравнения для однородного пространства без посторонних токов и зарядов являются плоские волны

$$\vec{C}(x,y,z,t) = \vec{C}\exp i(k_x x + k_y y + k_z z - \omega t), \; k_x^2 + k_y^2 + k_z^2 = (\omega/c)^2 \varepsilon\mu. \tag{1}$$

Квадраты компонент волнового вектора в (1) положительны: $k_x^2, k_y^2, k_z^2 > 0$. Однако решением волнового уравнения является и волна вида (1), но с отрицательным значением одного из квадратов компонент волнового вектора. Для определенности будем считать $k_z^2 < 0$ и $k_y^2 = 0$. При этом видно, что $k_x^2 > (\omega/c)^2 \varepsilon\mu$.

$$\vec{C}(x,y,z,t) = \vec{C}e^{-\kappa_z z}\exp i(k_x x - \omega t), \; \kappa_z = ik_z. \tag{2}$$

С физической точки зрения эта волна неприемлема, так как она неограниченно, экспоненциально нарастает в направлении отрицательных $z$. Если однако поместить в каком-либо сечении $z = const$ стенку, то бесконечность исчезнет, и решение окажется приемлемым и с физической точки зрения. Волну вида (2) называют неоднородной плоской волной. В отличие от (1) в волне (2) в плоскости постоянной фазы амплитуда волны не остается постоянной.

Волна вида (2) существует, например, у поверхности гипотенузной грани призмы полного внутреннего отражения. Поле за экраном со щелью в непосредственной близости к нему содержит неоднородные плоские волны наряду с однородными плоскими волнами. В неоднородном пространстве, содержащем слой вещества с $\varepsilon < 0$, на границе раздела сред существуют волны вида (2), называемые поверхностными поляритонами и поверхностными плазмонами. Окружающая среда имеет $\varepsilon \geq 1$. Магнитные проницаемости $\mu_i$ равны 1. Диэлектрическая проницаемость $\varepsilon < 0$ имеется у ионных кристаллов в ИК области, у хорошо отражающих свет металлов ($Ag, Au, Cu$) в оптической области.

Рассмотрим пространство с тонким слоем металла $\varepsilon < 0$, см. рис.1. Среди множества волн этого пространства в данный момент нас будут интересовать волны вида (2), то есть поверхностные волны, плазмоны. Оси $x$ и $y$ расположены в плоскости границы раздела, ось $z$ направлена по нормали к границе раздела.

Решение находят, задавая поля в виде (2) и рассматривая граничные условия на границах раздела сред. Граничные условия состоят в равенстве тангенциальных компонент полей $\vec{E}$ и $\vec{H}$ по обе стороны границы раздела. В результате возникают картины поля, изображенные на рис.2. При расчете картин поля приняты следующие значения входящих в расчет величин: $k_0 = 1.221\cdot 10^5\, см^{-1}$, $\varepsilon_1 = 1$, $\varepsilon_2 = -10.67$, толщина слоя серебра $d_2 - d_1 = d = 50\, nm$. Это поперечно-магнитные волны, у которых $H_x, H_z = 0$, $H_y \neq 0$, $E_x, E_z \neq 0$, $E_y = 0$. На каждой заданной частоте существуют две поверхностные волны, отличающиеся значением $k_x$, симметричный плазмон $k_{sx}$, фрагмент $a$ и антисимметричный плазмон $k_{ax}$, фрагмент $b$. Для выбранных значений параметров волновые числа плазмонов равны $k_{sx} = 1.259\cdot 10^5\, cm^{-1}$ и $k_{ax} = 1.321\cdot 10^5\, cm^{-1}$. Амплитуды поля измерены в единицах амплитуды возбуждающей волны.

Симметричный и антисимметричный плазмоны являются собственными волнами пространства со слоем металла. С помощью внешней прилегающей волны можно возбудить в слое металла поверхностную волну с произвольным значением волнового числа. Если волновое число возбуждающей волны близко к одному из резонансных значений, то есть либо к $k_{sx}$, либо $k_{ax}$, то амплитуда возбужденного в металле поля очень велика.

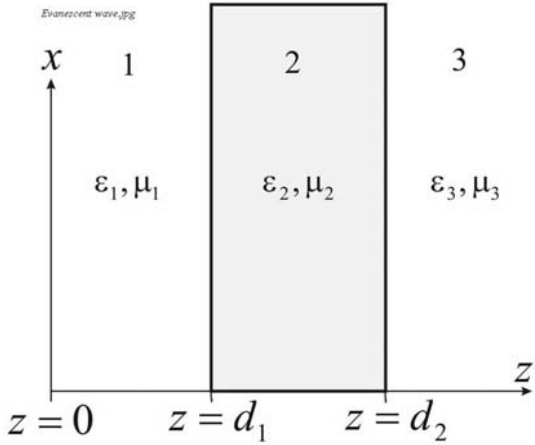
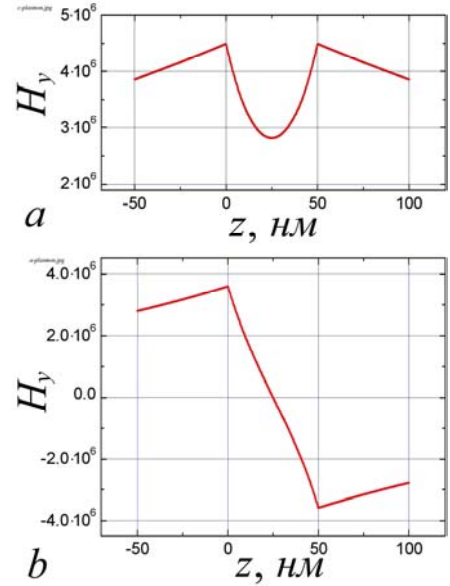

Рис.1. Пространство с плоским слоем металла, 1 и 3 - диэлектрик, $\varepsilon_1, \varepsilon_3 > 1$, 2 - металл, $\varepsilon_2 < 0$, $|\varepsilon_2| > \varepsilon_1, \varepsilon_3$.

Рис.2. $k_0 = 1.221 \cdot 10^5 \, см^{-1}$, $\varepsilon_1 = 1$, $\varepsilon_2 = -10.67$, толщина слоя серебра $d_2 - d_1 = 50 \, nm$,
$a$. симметричный плазмон $k_{sx} = 1.259 \cdot 10^5 \, cm^{-1}$,
$b$. антисимметричный плазмон $k_{ax} = 1.321 \cdot 10^5 \, cm^{-1}$.

При уменьшении толщины слоя металла $d$ различие между $k_{sx}$ и $k_{ax}$ растет. При $d = 5 \, nm$ и прежних значениях $k_0$, $\varepsilon_1$ и $\varepsilon_2$ волновые числа резонансных плазмонов равны $k_{sx} = 1.222 \cdot 10^5 \, cm^{-1}$ и $k_{ax} = 3.966 \cdot 10^5 \, cm^{-1}$. Различие между $k_{sx}$ и $k_{ax}$ увеличивается при уменьшении абсолютной величины $\varepsilon_2$. Так при $\varepsilon_2 = -1.2$, $\varepsilon_1 = 1$, $d = 50 \, nm$ и $k_0 = 1.221 \cdot 10^5 \, см^{-1}$ имеем $k_{sx} = 1.427 \cdot 10^5 \, cm^{-1}$ и $k_{ax} = 5.372 \cdot 10^5 \, cm^{-1}$. Впрочем после определенного выбора $\omega$, то есть $k_0$ не существует произвола в выборе $\varepsilon_2$, так как оно определяется физическими свойствами выбранного металла. Возможно, действующее значение $\varepsilon_2$ можно уменьшить, применяя перфорированные пленки металла. Вопрос требует специального изучения.

Можно изменять $\varepsilon_1$. При приближении значения $\varepsilon_1$ к $|\varepsilon_2|$, к значению модуля $\varepsilon_2$ растут значения $k_{sx}$ и $k_{ax}$, но относительная разница между ними не становится большой. Так при $\varepsilon_1 = 9$, $\varepsilon_2 = -10.67$, $d = 50 \, nm$ и $k_0 = 1.221 \cdot 10^5 \, см^{-1}$ получаем $k_{sx} = 8.867 \cdot 10^5 \, cm^{-1}$ и $k_{ax} = 9.577 \cdot 10^5 \, cm^{-1}$.

При увеличении толщины слоя металла расщепление на симметричный и антисимметричный плазмон исчезает, остается лишь один хорошо известный поверхностный плазмон $k_x = \frac{\omega}{c}\sqrt{\frac{\varepsilon_1 \varepsilon_2}{\varepsilon_1 + \varepsilon_2}}$. При $\varepsilon_1 = 1$, $\varepsilon_2 = -10.67$ и $k_0 = 1.221 \cdot 10^5 \, см^{-1}$ плазмон на одиночной границе имеет $k_x = 1.283 \cdot 10^5 \, cm^{-1}$, что не сильно отличается от $k_0$. При уменьшении $|\varepsilon_2|$ волновое число $k_x$ этого одиночного плазмона увеличивается.

Оба плазмона на тонком слое металла являются волнами более медленными, чем плоская волна в прилегающем пространстве. В наибольшей степени замедлен антисимметричный плазмон. Выше, в примере с пленкой толщиной $d = 5 \, nm$ его фазовая скорость в $3.27$ раза меньше, чем скорость плоской волны в прилегающем пространстве. Именно этот плазмон представляет интерес как очень медленный поверхностный плазмон.

Теперь рассмотрим плазмоны на металлическом наноцилиндре, см. рис.3. Поверхностная $TM_0$ волна существует на цилиндрах сколь угодно малого диаметра. Это поперечно-магнитная волна вида $\vec{C}(r, \theta, z, t) = \vec{C}(r) e^{i(hz - \omega t)}$, у которой $H_r, H_z = 0$, $H_\theta \neq 0$, $E_r, E_z \neq 0$, $E_\theta = 0$. На рис.4 показаны значения волнового числа в зависимости от радиуса наноцилиндра. Учтены потери в металле, поэтому волновое число – величина комплексная. Данные приведены в нормированном виде: $h_0 = h/k_0$, $a_0 = ak_0$, $k_0 = \omega/c$. Расчеты проделаны для наноцилиндра из серебра, длина волны $\lambda_0 = 514.5\ nm$ ($k_0 = 1.221 \cdot 10^5\ cm^{-1}$), $\varepsilon_2 = -10.67 + 0.326i$. Значение $a_0 = 0.01$ соответствует $a = 0.82\ nm$. При этом $Reh_0 = 38$, $Imh_0 = 0.842$. Длина волны плазмона равна $\lambda_{pl} = 13.5\ nm$, что в 38 раз меньше $\lambda_0 = 2\pi/k_0$, длина пробега плазмона $1/Imh_0$ составляет 45 длин волн плазмона.

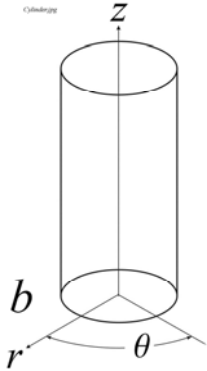

Рис.3. Пространство $\varepsilon_1 = 1$ с металлическим наноцилиндром $\varepsilon_2 < 0$, $|\varepsilon_2| > \varepsilon_1$.

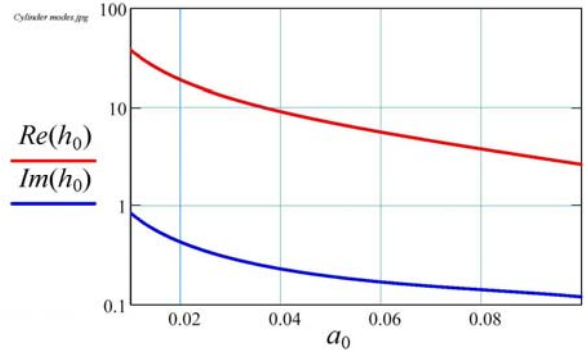

Рис.4. Наноцилиндр: значение волнового числа $h_0$ поверхностной $TM_0$ - волны в зависимости от радиуса $a_0$ цилиндра из $Ag$ на длине волны $514.5\ nm$, $k_0 = 1.221 \cdot 10^5\ cm^{-1}$.

Поле $TM_0$ - моды описывается следующими формулами:

$$\vec{E}_1 = \frac{\omega}{c}|A_2|R_A\left\{-\frac{h\gamma_1}{k_1}K_1(\gamma_1 r)\cos(hz-\omega t)\cdot\vec{i}_r + \frac{\gamma_1^2}{k_1}K_0(\gamma_1 r)\sin(hz-\omega t)\cdot\vec{i}_z\right\}$$

$$\vec{E}_2 = \frac{\omega}{c}|A_2|\left\{\frac{h\gamma_2}{\kappa_2}I_1(\gamma_2 r)\cos(hz-\omega t)\cdot\vec{i}_r + \frac{\gamma_2^2}{\kappa_2}I_0(\gamma_2 r)\sin(hz-\omega t)\cdot\vec{i}_z\right\}$$

$$\vec{H}_1 = |A_2|R_A\frac{k_1\gamma_1}{\mu_1}K_1(\gamma_1 r)\cos(hz-\omega t)\cdot\vec{i}_\theta$$

$$\vec{H}_2 = |A_2|\frac{k_2\gamma_2}{\mu_2}I_1(\gamma_2 r)\cos(hz-\omega t)\cdot\vec{i}_\theta$$

(3)

Магнитное поле находится в фазе (или в противофазе, что не существенно) с $r-$компонентой электрического поля и в квадратуре с $z-$компонентой. Вспомним, что в бегущей плоской волне свободного пространства магнитное поле и электрическое поле синфазны.

Теперь рассмотрим сферу и сфероид из металла $\varepsilon_2 < 0$ в пространстве $\varepsilon_1 = 1, |\varepsilon_2| > \varepsilon_1$. Для описания полей в этих частицах будем применять не традиционный метод, который в данной случае оказывается не наглядным, а иной, приближенный. Будем считать, что сфера и сфероид являются отрезками наноцилиндра. Оказывается, что для наносферы с $\varepsilon_2 = -2$ резонанс на $TM_0$ - моде случается приблизительно при длине эквивалентного наноцилиндра, равной $2a$. А это и есть сфера. Этот результат совпадает (приближенно, но с большой точностью) с тем результатом, который следует из стандартного строгого рассмотрения сферических волн в пространстве с металлической сферой. Для наносфероида с заданным соотношением полуосей $c/a$ резонанс по приближенному расчету, в котором сфероид заменяется отрезком наноцилиндра, случается при значении $\varepsilon_2$ (приближенно, но с убедительной точностью), которое соответствует значению $\varepsilon_2$, определяемому из точного расчета с разложением по сфероидальным функциям.

Для дальнейшего полезно отметить следующие обстоятельства. Пространство, полностью заполненное металлом, является 3-хмерным однородным пространством, пространство с плоским нанослоем металла является 2-мерным однородным пространством, пространство с металлической нанонитью является 1-мерным однородным пространством. Пространства со сферой и наносфероидом также являются 1-мерным пространством, но неоднородным. Собственными волнами наночастиц являются стоячие волны. В однородных пространствах имеют место определенные законы сохранения. В дальнейшем нас будет интересовать закон сохранения импульса (количества движения) соответствующих волн, двумерного импульса – в плоском слое, одномерного – в нити. Что касается сферы и сфероида, то особенность в том, что в них собственной волной является стоячая волна, а стоячие волны не переносят импульса. Стоячие волны образованы двумя встречными волнами, импульсы которых при сложении уничтожаются.

### *Спонтанное излучение атома вблизи наносферы из серебра*

Представляет интерес спонтанное излучение возбужденных атомов (молекул) вблизи металлической наночастицы. Опыты и расчет обнаруживают сильное возрастание интенсивности спонтанного излучения атомов вблизи наночастиц.

Интенсивность спонтанного излучения атома определяется квадратом напряженности поля моды пространства, в которую происходит испускание фотона, и числом эквивалентных мод поля (числом радиационных осцилляторов поля):

$$I_{sp} \sim (d \cdot E)^2 \rho_{E_n} \qquad (4)$$

Моды поля в пространстве с нанотелом отличаются от мод свободного пространства. В свободном пространстве напряженность поля моды, а эта мода представляет собой плоскую волну, всюду одинакова, в пространстве с наносферой поле моды пространственно неоднородно. Вблизи нанотела поле моды, как мы увидим ниже, сильно увеличено по сравнению с полем моды свободного пространства. Оказывается иным и число радиационных осцилляторов поля.

На увеличение вероятности спонтанного излучения атома в неоднородном пространстве благодаря увеличению числа эквивалентных радиационных осцилляторов поля обратил в свое время внимание известный Ю. Персел /12/.

В свободном пространстве набор осцилляторов поля можно выбрать произвольно. Это либо плоские волны в декартовых координатах, либо сферические волны в сферических координатах и т. д. И для неоднородного пространства выбор осцилляторов поля также может быть произвольным. Осцилляторы поля образуют полную систему функций, с помощью которых можно представить любое физическое поле. Целесообразно делать такой выбор, при котором поверхность раздела сред совпадает с какой-либо координатной поверхностью.

Рассмотрим наносферу. Для сферы целесообразно выбирать сферические координаты и моды поля в виде сферических волн. Нас будет интересовать мода $TM_{10}$. Это электродипольная мода. Схематически поле этой моды в свободном пространстве и в пространстве с наносферой показано на рис.5 *a* и *b*.

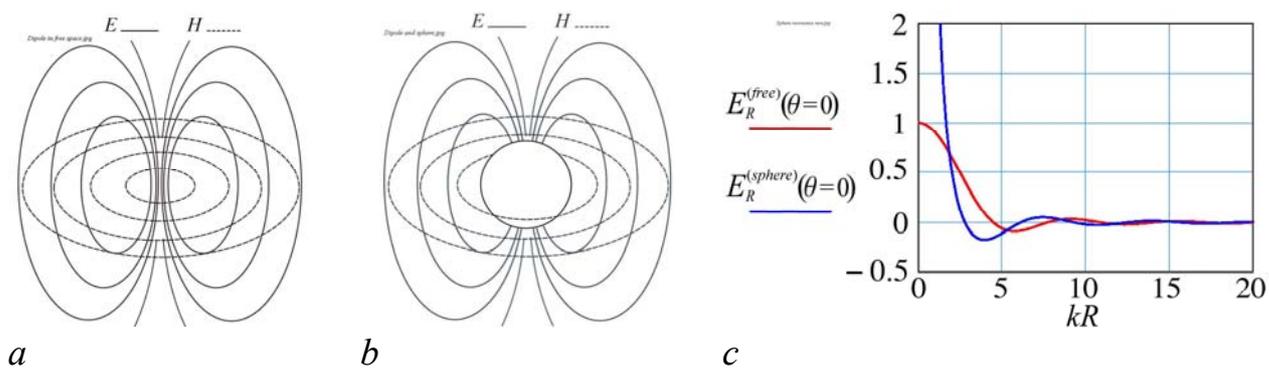

*a*          *b*          *c*

Рис.5. *a* – мода $TM_{10}$ свободного пространства, *b* – мода $TM_{10}$ пространства с металлической сферой, *c* – электрическое поле $TM_{10}$ в зависимости

В свободном пространстве поле этой моды пропорционально $j_1(k_1 R)$. На рис.5 *c* эта функция изображена красной кривой. Функция $j_1(k_1 R)$ ограничена в нуле. В пространстве с наносферой поле (вне сферы) пропорционально $y_1(k_1 R)$, на рис.5 *c* изображена синей кривой. На значительном удалении от начала координат $j_1(k_1 R) \sim \dfrac{\sin k_1 R}{k_1 R}$, а $y_1(k_1 R) \sim \dfrac{\cos k_1 R}{k_1 R}$. При $k_1 R \gg 1$ эти функции по амплитуде совпадают.

Вблизи начала координат функция $j_1(k_1R)$ ограничена, а $y_1(k_1R)$ нарастает неограниченно. Этими функциями описывается $R$ - компонента электрического поля $TM_{10}$ - моды. Обозначим радиус наносферы с помощью $R = a$. У поверхности наносферы $R$ - компонента электрического поля в $\left|\dfrac{y_1(k_1a)}{j_1(k_1a)}\right| \approx \dfrac{2}{(k_1a)^3}$ раз превышает поле $TM_{10}$ - моды в свободном пространстве. При $a = 10\ nm$ и $k_1 = 1.5 \cdot 10^5\ cm^{-1}$ превышение оказывается равным $592,5$. Вероятность спонтанного излучения пропорционально квадрату этого числа, то есть $3.5 \cdot 10^5$. В результате время спонтанного излучения сокращается от $10^{-8}\ c$ до $3 \cdot 10^{-14}\ c$. Сечение комбинационного рассеяния, пропорциональное 4-й степени поля и исходно равное, скажем, $10^{-30}\ см^2$, увеличивается в $1.2 \cdot 10^{11}$ раз. Это означает, что вероятность комбинационного рассеяния оказывается равной вероятности спонтанного излучения, что и наблюдали Фелд и коллеги /13/.

Резонанс наносферы наблюдается при $\varepsilon_2 = -2$, то есть при определенной частоте. Резонанс в наносфероиде может наблюдаться при иных значениях $\varepsilon_2$, что определяется соотношением осей сфероида.

### *Фокусировка оптического излучения в пятно диаметром меньше длины волны*

Яблонович в докладах с названием "Plasmon wave imaging: Optical frequencies but with X-ray wavelengths!", читанных в 2007 г., предлагал опыт, схема которого изображена на рис.6.

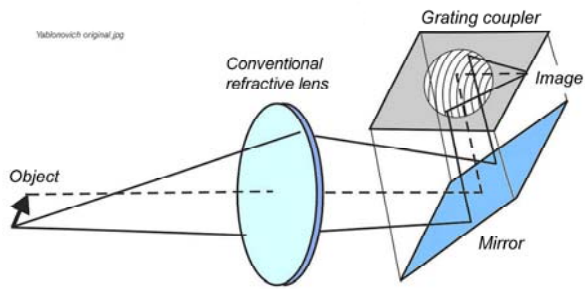  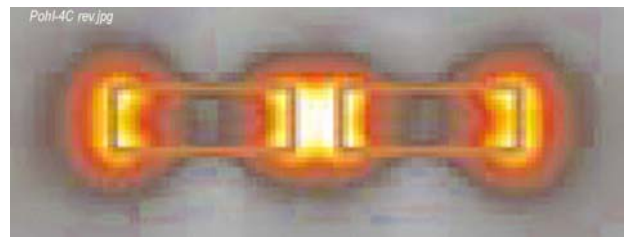

Рис.6.  Рис.7.

Объект фокусируется с помощью обычной линзы на поверхность тонкой пленки хорошо отражающего свет металла. В пятне фокусировки имеются штрихи в виде отрезков окружностей с общим центром. Эти штрихи образуют дифракционную решетку, с помощью которой на пленке возбуждается сходящийся пучок поверхностных волн – плазмонов. В точке схождения плазмонов образуется пятно, размеры которого определяются длиной волны плазмона, то есть пятно меньше традиционного дифракционного, которое имеет размер порядка $\lambda_0$. Осуществляется субволновая фокусировка света.

Размер фокусного пятна в сходящемся пучке плазмонов определим с помощью представления о гауссовом пучке. Гауссов пучок оптических плазмонов описывается формулой

$$u(x, y, z) = \sqrt{-i\dfrac{ha^2}{x - iha^2}} \exp\left(\dfrac{i}{2}\dfrac{hy^2}{x - iha^2} - qz + ihx\right). \qquad (5)$$

Оси координат $x$ и $y$ находятся в плоскости пленки, ось $z$ направлена по нормали к пленке. Эта формула совпадает с формулой для двумерного гауссова пучка в свободном пространстве за исключением того, что вместо $k = (\omega/c)(\varepsilon\mu)^{1/2}$ в формуле (5) фигурирует волновое число плазмона $h$, которое много больше, чем $(\omega/c)(\varepsilon\mu)^{1/2}$, и экспонента $e^{-qz}$, которая ($q$ принимает 4 значения, по два значения для каждой границы раздела) описывает экспоненциальное убывание поля поверхностного плазмона при удалении от поверхности. Формула (5) выведена при условии $ha \gg 1$, которое сходно с обычным условием для гауссова пучка $ka \gg 1$.

С помощью представления о гауссовом пучке плазмонов возникает оценка для наименьшего поперечного размера сходящегося пучка плазмонов (для так называемой перетяжки) $2a \gg \lambda_{pl}/\pi$. Можно считать, что $(2a)_{min} = 3\lambda_{pl}$. Поперечный размер в перпендикулярном к пленке толщины $d$ направлении равен $1/q_1 + 1/q_3 + t \approx 2/h + d = \lambda_{pl}/\pi + d$.

Проделаем расчеты для пленки серебра толщиной $1\ nm$ и длины волны $\lambda_0 = 514.6\ nm$. В среде $\varepsilon_{1,3} = 1$ для пленки $\varepsilon_2 = -10.67$ /8/ ($\mu_{i=1,2,3} = 1$) при $k_0 = 2\pi/\lambda_0 = 1.221 \cdot 10^5\ cm^{-1}$ резонансное волновое число $h = \sqrt{q_2^2 + (\omega/c)^2 \varepsilon_2 \mu_2}$ медленного, антисимметричного плазмона оказывается равным $1.88 \cdot 10^6\ cm^{-1}$, что в 15.4 раза больше, чем $k_0$. Если взять $\varepsilon_1 = \varepsilon_3 = 6.5$, то $h$ оказывается равным $1.42 \cdot 10^7\ cm^{-1}$, что уже в 116 раз больше $k_0$. В 1-ом случае $\lambda_{pl} = 33.4\ nm$, $q_{1,2} \approx 1.9 \cdot 10^6\ cm^{-1}$, во 2-м случае $\lambda_{pl} = 4.44\ nm$, $q_{1,2} \approx 1.4 \cdot 10^7\ cm^{-1}$. Размер перетяжки гауссова пучка плазмонов в 1-м случае равен $3\lambda_{pl} \times (\lambda_{pl}/\pi + d) = 100$ нм $\times$ 12 нм, во 2-м случае $3\lambda_{pl} \times (\lambda_{pl}/\pi + d) = 13.5$ нм $\times$ 2.4 нм.

Мы, действительно, обнаруживаем возможность произвести субволновую фокусировку света. И этим объясняется энтузиазм, который существует по поводу фокусировки плазмонов. К сожалению, имеется одно осложняющее обстоятельство. Это потери распространения. Гауссов пучок плазмонов в рассматриваемом устройстве имеет довольно большие размеры, будучи измеренным в длинах волн плазмонов: 30-100 волн по фронту в начальной точке и 300-1000 волн вдоль оси.

Вычисления дают следующий результат. При $\lambda_0 = 514.6\ nm$, $\varepsilon_1 = \varepsilon_3 = 1$, $\varepsilon_2 = -10.67 + i0.3$ пробег плазмона, который равен отношению $Re h / Im h$, составляет 35.5 волн. При $\varepsilon_1 = \varepsilon_3 = 6.5$ пробег плазмона составляет 26 волн. При $\varepsilon_1 = \varepsilon_3 = 10$ длина волны плазмона в 273.5 раза короче длины волны в вакууме на той же частоте, и пробег плазмона составляет всего 8.2 волн.

Большие потери распространения являются серьезной проблемой для применений. Однако имеется возможность применить прилегающие усиливающие слои $\varepsilon_{1,3} = \varepsilon_{1,3}^{'} - i\varepsilon_{1,3}^{''}$ для компенсации потерь распространения. Имеются сообщения о таких успешных опытах /14/. Расчет показывает, что при $\varepsilon_1 = 9 - i0.27$ оказывается, что $h = (147.4 + i5.3 \cdot 10^{-3})(\omega_0/c)$. Для предотвращения усиления мало замедленного симметричного плазмона усиливающий слой следует брать тонким.

Вопрос о потерях является решающим, если иметь в виду импульсы большой мощности. Сквозь плазмонный волновод с усиливающим слоем может пройти лишь импульс, не насыщающий усиливающий слой. Интенсивность насыщения $I_s$ равна $\hbar\omega_0/\sigma\tau$. Если сечение усиления $\sigma$ взять равным $2.5 \cdot 10^{-20}\ cm^2$ и время релаксации возбуждения $\tau$ равным $10^{-10}\ s$, то при $\lambda_0 = 514.6\ nm$ получаем $I_s = 1.5 \cdot 10^{11}\ watt/cm^2$. Это довольно умеренное значение интенсивности. Если иметь в виду импульс длительности 100 фемтосек, то вместо $\tau$ следует взять длительность импульса. В результате получим $I_s = 1.5 \cdot 10^{14}\ watt/cm^2$. Впрочем, такую и даже заметно больше интенсивность можно получить и без субволной фокусировки.

Вопрос об интенсивности насыщения не возникает, если плазмонный волновод применять в приемнике пороговой чувствительности. В приемнике необходимо применять устройство того или иного вида для пространственной селекции мод.

Теперь рассмотрим второе устройство, которую назовем резонансной оптической антенной Поля, рис.7. В устройстве Поля иследовалась полоска из золота толщиной 40 нм, длиной ~ 260 нм и шириной 45 нм с разрезом посередине шириной ~ 20 нм.

При рассмотрении этой антенны возникает вопрос о том, в какой мере она перекрывает пятно фокусировки диаметра $\lambda_0$, то есть вопрос о сечении рассеяния.

Проверяем, каковы должны быть по теории размеры такой полоски в условиях резонанса. Расчет проделываем приближенно, заменяя полоску отрезком наноцилиндра и пользуясь характеристическим уравнением для этого случая. При диаметре наноцилиндра $42\ nm$ резонансная длина по расчету оказывается равной $L = 244.6\ nm$. Это значение хорошо согласуется с указанным в статье Поля значением ~ 260 нм.

Напряженность поля $\vec{E}$ в зазоре между половинками антенны равна электрической индукции $\vec{D}$ в материале антенны. Это утверждение возникает в результате использования граничного условия о равенстве нормальных компонент электрической индукции по обе стороны границы раздела двух сред с разным $\varepsilon$.

Для простоты рассмотрения заменим полоску вытянутым металлической сфероидом объема $V$ с полуосями $a$ и $c$ во внешнем электромагнитном поле $E_0$ частоты $\omega_0$. Частица имеет $\varepsilon_1(\omega) = \varepsilon_1'(\omega) + i\varepsilon_1''(\omega)$. Внешнее электрическое поле направлено вдоль самой длинной, главной оси сфероида. Мы будем пользоваться формулами из работы /15/, где рассмотрен резонанс в наносфероиде. Полный дипольный момент, наведенный в сфероиде, описывается формулой

$$P(\omega_0) = \chi(\omega_0)V[E_0 + i(2/3)k_0^3 P(\omega_0)], \tag{6}$$

где $\chi(\omega)$ - диагональный элемент тензора восприимчивости сфероида. При $V/\lambda^3 \ll 1$ величина $\chi(\omega)$ может быть вычислена в электростатическом приближении. Второй член в (6) учитывает энергетические потери из-за излучения посредством введения реакции поля излучения $\vec{E}_r = (2/3)k_0^3 \vec{P}(\omega)$, $k_0 = \omega_0/c$. Поле $\vec{E}_r$ выбрано так, как это делают в классической теории поля: работа, которую производит $\vec{E}_r$ над диполем $\vec{P}$, равна излученной энергии. Разрешив уравнение (6) относительно $P(\omega_0)$ получаем:

$$P(\omega_0) = \frac{\chi(\omega_0)}{1 - i(2/3)k_0^3 \chi(\omega_0)V} V E_0. \tag{7}$$

Поправочный член в знаменателе (7) имеет сильную зависимость от частоты из-за $k_0^3$ и $\chi(\omega_0)$. Последняя величина становится большой при плазмонных резонансах в частице. Для восприимчивости сфероидальной частицы имеет место формула

$$\chi(\omega) = \frac{1}{4\pi} \frac{\varepsilon_1(\omega) - 1}{2 - [1 - \varepsilon_1(\omega)]A}. \tag{8}$$

Фактор деполяризации $0 < A < 1$ характеризует эксцентричность частицы.

Для золота $\varepsilon_1''$ мало в видимом спектре. Поэтому пиковое возрастание находим, максимизируя (8) по отношению к $\varepsilon_1'(\omega)$. Для $\varepsilon_1''$, $V/\lambda^3 \ll 1$ максимум достигается при $2 - [1 - \varepsilon_1'(\omega_{res})]A \approx 0$. Использовав это условие в (8) мы видим, что радиационное затухание следует принимать в расчет всякий раз, когда $(1 - \varepsilon_1') \times (4\pi^2 V / 3\lambda^3) \approx \varepsilon_1'' A$.

Получим формулу для $P$ в максимуме. Подставим в (7) резонансное $\varepsilon_1'$, определяемое условием $2 - [1 - \varepsilon_1'(\omega_{res})]A \approx 0$.

$$P_{res} = -i(3/2k_0^3) \frac{E_0}{[3\lambda^3/(2\pi)^2 V][\varepsilon_1'' A/(\varepsilon_1 - 1)] - 1}. \tag{9}$$

Внешнее поле направлено вдоль самой длинной оси сфероида.

Рассчитаем поле в зазоре. Оно равно $\vec{E}_{gap} = \vec{D} = \vec{E}_0(1 + 4\pi P_{res}/V)$.

$$\left|\vec{E}_{gap}/\vec{E}\right| \approx \{[\varepsilon_1'' A/(\varepsilon_1' - 1) - 4\pi^2 a^3 f / 3\lambda^3]\}^{-1}, \quad f = c/a. \tag{10}$$

Вычислим $\left|\vec{E}_{gap}/\vec{E}\right|$ для параметров опыта в /3/: $a = 21.12\ nm$, $c = 122.3\ nm$, $A = 0.091$, $f = 5.791$, $\lambda_0 = 830\ nm$, $\varepsilon_1 = -25.8 + i1.627$. Расчет дает $\left|\vec{E}_{gap}/\vec{E}\right| = 168.8$. На рис.8 приведена зависимость $\left|\vec{E}_{gap}/\vec{E}\right|$ от $a$ для золота на длине волны $\lambda_0 = 830\ nm$.

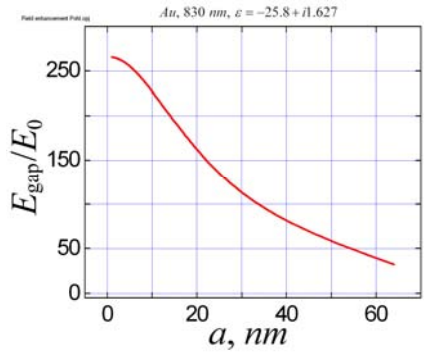
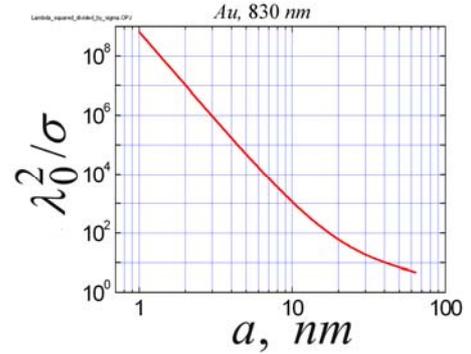

Рис.8.  Рис.9.

Рассчитаем сечение рассеяния, которое в соответствии с определением равно $\sigma = \dfrac{\langle W \rangle}{cE_0^2/8\pi}$. Здесь $\langle W \rangle$ - среднее значение мощности излучения диполя, $\langle W \rangle = ck_0^4 d_0^2/2$, $d_0$ - амплитуда осцилляций диполя, равная в данном случае $P_{res}$:

$$\langle W \rangle = \frac{3cE_0^2 \lambda^2}{16\pi^2 [3\varepsilon_1'' \lambda^3 A/(2\pi)^2 V(\varepsilon_1-1)-1]^2}. \tag{11}$$

Поэтому

$$\frac{\lambda^2}{\sigma} = \left[\frac{2\sqrt{\pi}\lambda^3 \varepsilon_1'' A}{(2\pi)^2 V(\varepsilon_1-1)} - \frac{2\sqrt{\pi}}{3}\right]^2, \tag{12}$$

$$A = \int_0^\infty a^2 c(s+c^2)^{-3/2}(s+a^2)^{-1} ds, \quad f = c/a \text{ - эксцентриситет.}$$

Зависимость отношения $\lambda_0^2/\sigma$ от $a$, радиуса нанопроволочки из золота в вакууме на длине волны $\lambda_0 = 830$ nm изображена на рис.9. При $a = 21.12$ nm отношение $\lambda_0^2/\sigma$ равно ~ 40. При $a = 128$ nm отношение $\lambda_0^2/\sigma$ равно ~ 2.7.

Из проделанного рассмотрения следует вывод: как устройство Яблоновича, так и устройство Поля, каждое способное локализовать электромагнитное поле на площадке, которая много меньше $\lambda^2$, не имеет 100%-ной эффективности.

Рассмотрим усовершенствование, которое пригодно как для устройства Яблоновича, так и для устройства Поля. Возьмем резонансный атом (квантовую точку) и поместим его вблизи тонкой пленки или сфероида. Будем возбуждать атом, а он, излучая, в свою очередь будет возбуждать наноустройство. Сечение поглощения атома равно $\cong \lambda_0^2 \Delta\omega_{rad}/\Delta\omega_{inh}$. Здесь $\Delta\omega_{rad}$ - радиационная ширина, $\Delta\omega_{inh}$ - допплеровская ширина в случае атома или ширина из-за взаимодействия с фононами в случае квантовой точки. В видимом спектре для атома $\Delta\omega_{rad}/\Delta\omega_{inh} \approx 0.01$, для квантовой точки $\Delta\omega_{rad}/\Delta\omega_{inh}$ может достигать значения $0.1$. Видно, что возбуждение атома и, в особенности, квантовой точки происходит с довольно высокой эффективностью.

Атом вблизи наносфероида, см. рис.10, находится в объеме двух мод неоднородного пространства. Первая мода – сферическая мода $\vec{n}_{10}$, возбуждающая волна. Вторая мода – резонансный поверхностный плазмон наносфероида $TM_0$. Падающая волна возбуждает атом и практически не возбуждает плазмон $TM_0$. Но спонтанный излучательный переход атома происходит преимущественно с излучением фотона в $TM_0$ плазмон. Эффективность возбуждения $TM_0$ плазмона равна практически 100%.

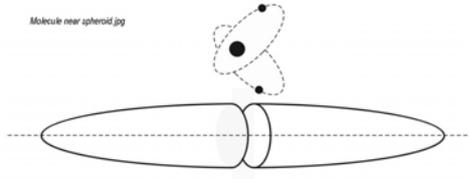

Рис.10.

Отношение вероятности излучения в плазмон к вероятности излучения в свободное пространство равно

$$F = E_{pl}^2 \rho_{pl} / E_{free}^2 \rho_{free}, \qquad (13)$$

где $E_{pl}$, $E_{free}$ - поле соответствующей моды поля, нормированное на энергию, равную одному фотону, $\rho_{pl}$, $\rho_{free}$ - плотность состояний соответствующего поля. Вычисления показывают, что при максимальной связи атома с плазмоном

$$F \approx (h/k_0)^3 \varepsilon_1^2 Q, \qquad (14)$$

где $h$ - волновое число плазмона, $Q$ - добротность плазмонного резонанса в наносфероиде длиной $L = \pi/h$, $k_0 = \omega/c$. Расчет показывает, что для наносфероида диаметром $2\ nm$ отношение $h/k_0 \approx 33$ и $Q \approx 70$ (определяется потерями в веществе наносфероида, радиационные потери малы) для $\lambda_0 = 514.6\ nm$, $\varepsilon_1 = -10 + i0.3$ ($Ag$). При этих значениях входящих в выражение для $F$ величин оно оказывается равным $3 \cdot 10^8$. Действительно, возбужденный атом излучает преимущественно в плазмон. Полученную величину $F$ следует уточнить, не пользуясь приемами теории возмущений. Однако она и в этом случае будет чрезвычайно большой.

При большом значении $F$ оказывается большой величина $\Delta\omega_{inh}$, которая теперь равна обратному времени излучения в плазмон, то есть равна $\Delta\omega_{rad}F$. Связь с плазмоном следует уменьшить до умеренно малой величины. Это может быть достигнуто при расположении атома на удалении от наносфероида. Выбрав $F$ равным $2 \div 3$ получим по-прежнему практически 100%- ную эффективность излучения в плазмон практически без увеличения отношения $\Delta\omega_{rad}/\Delta\omega_{inh}$.

Подведем итог. Рассмотрены два варианта устройств субволновой фокусировки света. Первое устройство представляет собой тонкую пленку хорошо отражающего свет металла, на которой возбуждается сходящийся пучок поверхностных плазмонов с волновым числом $h >> \omega_0/c$. Перетяжка этого пучка имеет сечение много меньше $\lambda_0^2$. Второе устройство представляет собой линейную антенну наноразмеров с разрезом посередине. В антенне возбуждается резонансный поверхностный плазмон. Поле в разрезе концентрируется на площадке много меньше $\lambda_0^2$. И в 1-м, и во 2-м случае наблюдается эффект значительной интенсификации поля в пятне субволновой фокусировки. Однако полная мощность в выходном пятне ослаблена в 1-м случае в силу потерь распространения, в 2-м случае - в силу малого сечения возбуждения наноантенны. Предложено улучшить эффект возбуждения пленки и наноантенны посредством возбуждения прилегающего атома или квантовой точки.

### *Почти идеальная линза Пендри: проецирующие свойства тонкого слоя металла*

Пендри показал, что при применении гипотетических или искусственных материалов с отрицательным показателем преломления $n = -1$ можно построить изображение объекта с разрешением, которое заметно превышает разрешение обычного объектива. Линза из такого материала восстанавливает не только фазы распространяющихся волн, но и амплитуды исчезающих волн (evanescent waves). Если иметь в виду практически доступные оптические материалы, то подобный эффект будет, как показал Пендри, приближенно наблюдаться в опытах с тонкой пленкой серебра на частоте, где $\varepsilon = -1$ (ультрафиолетовая область).

Будем называть описанное свойство реального металла эффектом почти идеальной линзы Пендри или просто эффектом почти идеальной линзы. Работа Пендри инициировала многочисленные последующие работы. Среди них отметим исключительно важную работу /16/, где экспериментально отчетливо продемонстрирован эффект увеличения качества мелких деталей в изображении в опыте с плоским слоем серебра.

Утверждают, что технологам эффект известен давно. Тонкий слой металла якобы применяют при изготовлении ценных бумаг.

Подробное рассмотрение показывает, что существование эффекта почти идеальной линзы Пендри не ограничено условием $\varepsilon = -1$. Снова рассмотрим пространство с тонким плоским слоем хорошо отражающего свет металла, см. рис.11.

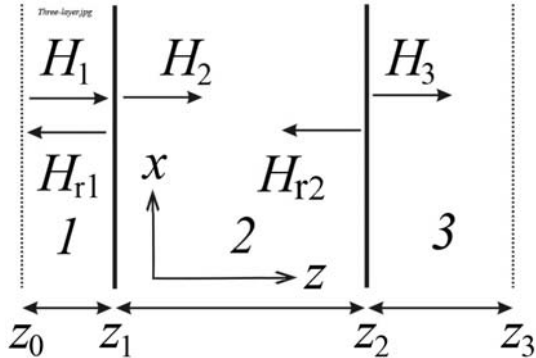 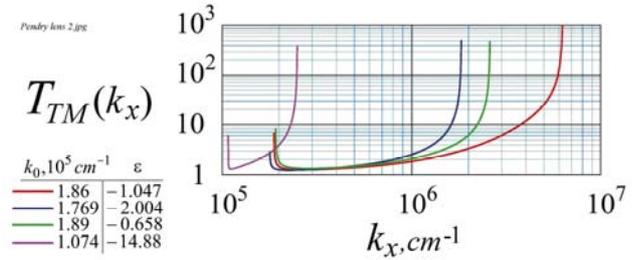

Рис.11. Пространство с тонким слоем металла и исчезающие волны в этом пространстве.   Рис.12.

В этом пространстве цифрами $1$ и $3$ обозначены промежутки без серебра, цифрой $2$ – слой серебра, а стрелками показаны убывающие волны в этом пространстве, которых всего пять. Начало координат находится на плоскости раздела сред $1$ и $2$. Стрелками показаны направления убывания амплитуд волн. В объектной плоскости $z_0 < 0$, параллельной слою металла, имеется источник в виде двух параллельных щелей. Картина поля в объектной плоскости имеет вид:

$$\vec{H}_1 = \vec{a}_y H_{1y} e^{-\kappa_{1z} z_0} \cdot f(x),$$
$$f(x) = \cos k_x x \cdot \{\exp[-(x-x_0)^m / \Delta x_0^m] + \exp[-(x+x_0)^m / \Delta x_0^m]\}, \quad (15)$$
$$m = 2^n, \quad n = 1,2..., \quad k_x^2 - \kappa_{1z}^2 = k_0^2 = (\omega/c)^2 \varepsilon_1 \mu_1.$$

Рассматриваются поперечно-магнитные волны, поляризованные по оси $y$. Суммарная толщина промежутков $z_0 z_1$ и $z_2 z_3$ вне серебра равна $L$. Пропускание слоя серебра найдем, составив граничные условия на каждой из границ раздела сред. Граничные условия заключаются в равенстве тангенциальных компонент полей $\vec{E}$ и $\vec{H}$ по обе стороны границ раздела. В результате возникает следующее выражение для функции пропускания слоя металла $T_{TM}(k_x)$:

$$T_{TM}(k_x) = H_{3y} e^{-\kappa_{1z} d} / H_{1y} = \frac{4K}{(K+1)^2 e^{-\kappa_{2z} d} - (K-1)^2 e^{\kappa_{2z} d}},$$
$$\kappa_{1z} = \sqrt{k_x^2 - (\omega_0/c)^2 \varepsilon_1 \mu_1}, \quad \kappa_{2z} = \sqrt{k_x^2 - (\omega_0/c)^2 \varepsilon_2 \mu_2}, \quad K = -\frac{\kappa_{2z} \varepsilon_1}{\kappa_{1z} \varepsilon_2}. \quad (16)$$

Это выражение совпадает с выражением для $T_p(k_x)$ из работы Пендри и соавторов /17/, если в последнем положить $k_z^{(1)} = i\kappa_{1z}$, $k_z^{(2)} = i\kappa_{2z}$, $\varepsilon_1 = \varepsilon_3$ и $\mu_1 = \mu_2 = \mu_3 = 1$. Ниже принято, что все магнитные проницаемости $\mu_i$ равны $1$.

На рис.12 приведено несколько кривых, изображающих функцию $T_{TM}(k_x)$ для слоя серебра толщиной $6$ $nm$. Кривые имеют особенности при значениях $k_x$, равных волновым числам двух резонансных поверхностных плазмонов, симметричного $k_s$ и антисимметричного $k_a$. В промежутке $(k_s, k_a)$ между волновыми числами резонансных плазмонов функция $T_{TM}(k_x)$ превышает единицу. Это означает, что за слоем металла поле усилено по сравнению с полем на ближней к объекту границе слоя металла. Усиление тем больше, чем ближе $k_x$ к значению $k_a$, наибольшего из резонансных волновых чисел. Заметим, что формула (16) для пропускания слоя металла представляет собой отношение амплитуды поля прошедшей волны к амплитуде поля падающей волны, а не к амплитуде суммарного поля падающей и отраженной волн на передней границе слоя металла.

На рис.12 можно видеть, как изменяется функция пропускания при изменении $\varepsilon$. При значении $\varepsilon$, близком к $-1$, расстояние между волновыми числами $k_s$ и $k_a$ резонансных плазмонов оказывается наибольшим. Число парциальных волн, подверженных усилению, нарастает при $\varepsilon \to -1$ Это означает повышение разрешающей способности проецирующей пленки серебра.

Поле по формуле (11) и спектр волновых чисел $Gt(k_x) = \dfrac{1}{\sqrt{2\pi}} \int\limits_{-\infty}^{\infty} f(x) e^{-ik_x x} dx$ составляющих его исчезающих волн $e^{ik_x x - \kappa_{1z} z}$ изображены на рис.13 $a$ и $b$.

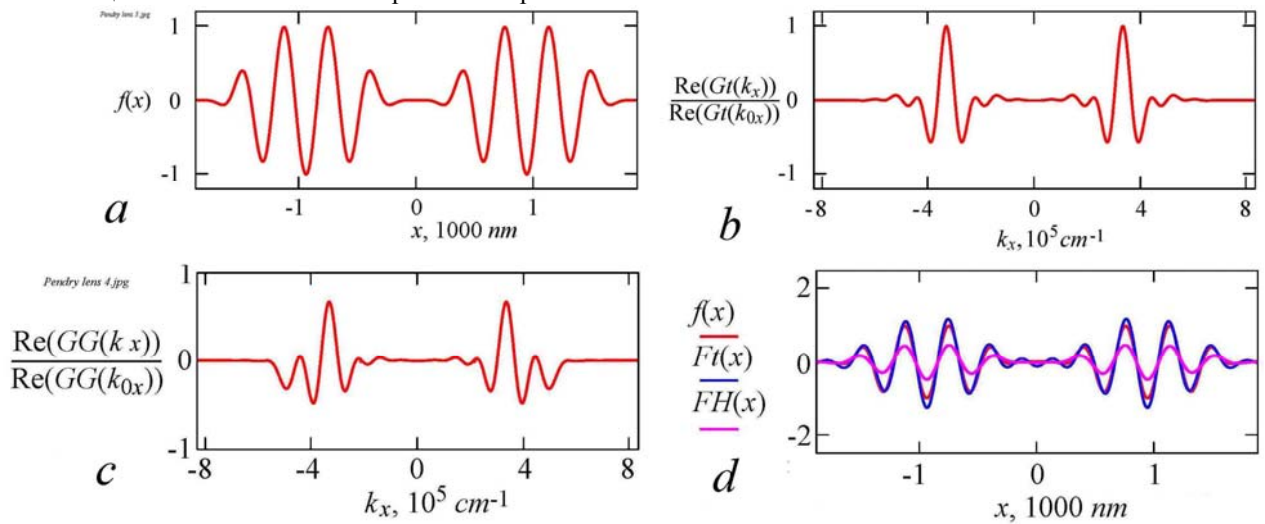

Рис.13.

Волновое число $k_{0x}$ выбрано в промежутке между $k_s$ и $k_a$. Оно равно $k_a/1.5 = 1.67 \cdot 10^5 \, см^{-1}$, длина волны $\lambda_0 = 376.2 \, нм$. Расстояние между серединами щелей $2x_0 = 2 \cdot 2.5 \cdot \lambda_0 = 1881 \, нм$, полная ширина каждой из щелей $2\Delta x_0 = 3 \cdot \lambda_0 = 1129 \, нм$. Толщина слоя серебра $d = 6 \, нм$. Суммарная толщина $L$ промежутков вне слоя серебра равна $10d = 60 \, нм$. На отрезке от $z_0$ до $z_3$ компоненты поля $e^{ik_x x - \kappa_{1z} z}$ подвергаются трансформации по формуле

$$GG(k_x) = Gt(k_x) T_{TM}(k_x) e^{-\kappa_{1z}(k_x) L}. \qquad (17)$$

Вычисленный по формуле (17) спектр изображен на рис.13 $c$. Из сравнения с кривой на рис.13 $b$ видны возникшие искажения при больших значениях $|k_x|$. Поле в плоскости $z_3$ получаем по формуле обратного фурье-преобразования

$$Ft(x) = \dfrac{1}{\sqrt{2\pi}} \int\limits_{0.7 k_{0x}}^{1.3 k_{0x}} Gt(k_x) T_{TM}(k_x) e^{-\kappa_{1z}(k_x) L} (e^{ik_x x} + e^{-ik_x x}) dk_x. \qquad (18)$$

Интегрирование в (18 4) проделывается в несколько ограниченном интервале значений $k_x$, что означает фильтрование. Результат вычислений по формуле (18) представлен синей кривой на рис.13 $d$. Это поле в плоскости изображений $z_3$. Поле в плоскости $z_0$ на этом рисунке изображено красной кривой. Пурпурным цветом изображено поле в плоскости $z_3$ в отсутствие слоя серебра. Для этого случая кривая рассчитана по формуле

$$FH(x) = \frac{1}{\sqrt{2\pi}} \int\limits_{0.7k_{0x}}^{1.3k_{0x}} Gt(k_x) e^{-\kappa_{1z}(k_x)(L+d)} (e^{ik_x x} + e^{-ik_x x}) dk_x. \quad (19)$$

Видно, что поле в плоскости $z_3$ в опыте со слоем серебра довольно хорошо совпадает с полем в объектной плоскости $z_0$. Наблюдаются некоторые искажения, но они невелики. Отметим, что плоскость $z_3$ находится на удалении $L + d = 66 нм$ от объектной плоскости. В опыте без слоя серебра поле в этой плоскости оказывается двукратно ослабленным, пурпурная кривая.

Теперь обсудим случай близких по абсолютной величине значений $\varepsilon_1$ и $\varepsilon_2$. На длине волны $430.8 нм$ для серебра $\varepsilon_2 = -6.06 + i0.197$, для алмаза - $\varepsilon_1 = 6.042$, отношение $\operatorname{Re}\varepsilon_2/\varepsilon_1$ равно $-1.003$. Ниже в расчете отношение $\varepsilon_2/\varepsilon_1$ равно $-1.01$. Для $\varepsilon_3$ выбираем значение, равное $\varepsilon_1$. Первоначально считаем $Im\varepsilon_2 = 0$. Толщина слоя серебра $d = 2 нм$. Суммарная толщина $L$ промежутков вне слоя серебра равна $d$. Соответствующие волновые числа равны: $k_0 = 1.458 \cdot 10^5 см^{-1}$, $k_s = 1.459 \cdot 10^5 см^{-1}$ и $k_a = 2.653 \cdot 10^7 см^{-1}$. Волновое число $k_{0x}$ выбрано равным $\frac{k_a}{1.5} = 1.769 \cdot 10^7 см^{-1}$.

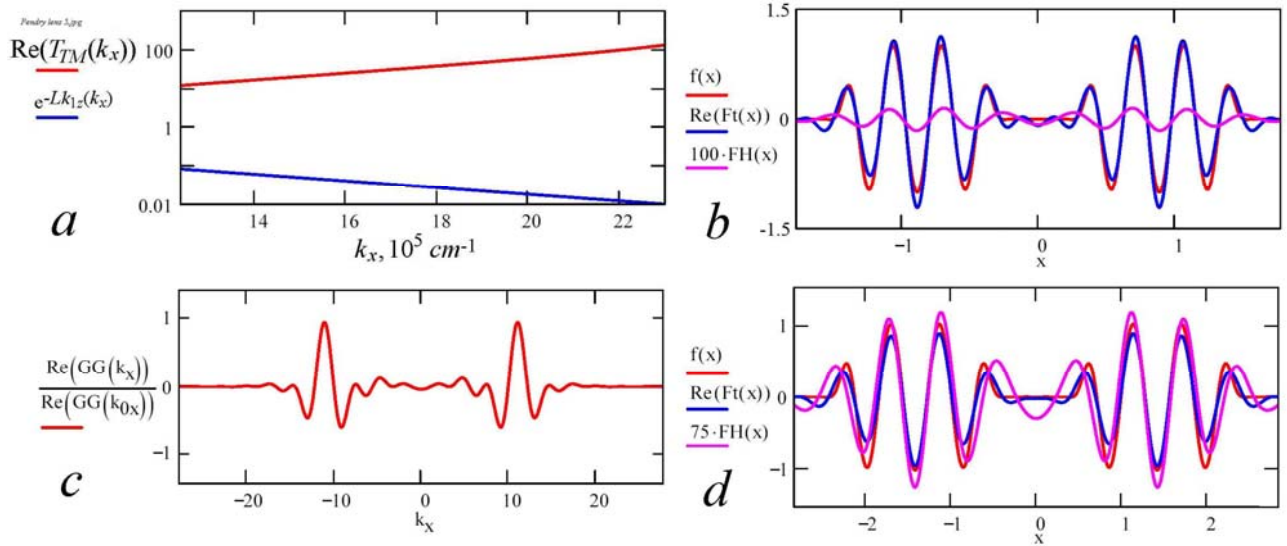

Рис.14.

На рис.14 *a* приведены функция $T_{TM}(k_x)$, верхняя, красная кривая, и функция $e^{-\kappa_{1z}L}$, нижняя, синяя кривая, в интервале значений $k_x = (0.7 \div 1.3)k_{0x}$. По оси ординат масштаб логарифмический. Функция $e^{-\kappa_{1z}L}$ принимает значения от $\sim 0.1$ до $0.01$. Соответственно пропускание слоя толщиной $L + d = 2d$, то есть квадрат этой функции, меняется в пределах от $\sim 0.01$ до $\sim 0.0001$. В опыте без слоя серебра сигнал в плоскости $z_3$ оказывается очень слабым. На рис.14 *b* этот сигнал, пурпурная кривая, изображен 100-кратно увеличенным. Кроме того видно, что сигнал сильно искажен. Промежуток между импульсами практически исчез.

Как и в предыдущем случае в опыте со слоем серебра в вакууме наблюдается хорошая передача сигнала из плоскости $z_0$ в плоскость $z_3$. Исходный сигнал изображен красной кривой, итоговый сигнал изображен синей кривой. Оба сигнала мало отличаются. Наблюдаются небольшие искажения в виде пульсаций в промежутке между импульсами. Однако контраст довольно велик и, соответственно, пространственное разрешение можно признать хорошим.

Теперь учтем потери в металле. Для диэлектрической проницаемости металла выберем значение $\varepsilon_2 = -6.06 + i0.197$, для прилегающего слоя - $\varepsilon_1 = 6.042$. Остальные параметры – как для рис.14 *a* и *b*.

Результат расчета приведен на рис.14 *c* и *d*. Отличия заметны лишь на кривой $GG(k_x)$. Функция $Ft(x)$ с потерями, рис.14 *d*, практически не отличима от таковой без потерь, рис.14 *b*. Таким образом, можно признать, что при выбранных значениях параметров влияние потерь незначительно. Этот результат отличается от результата Пендри, представленного в /16/.

Выбранные выше значения толщины слоя серебра довольно малы, и возникает вопрос о применимости выбора для $\varepsilon_2$ значений, определенных для массивных образцов серебра. Согласно /18/ в наночастицах (кластерах) атомов хороших металлов диаметра $2R = 10\,нм$ длина свободного пробега электронов заметно сокращается из-за возрастающей роли столкновений с поверхностью. Для серебра длина свободного пробега $l_\infty$ электронов в массивном образце равна $52\,K$ при $273\,K$ /18/. Уменьшение длины пробега приводит к изменению диэлектрической проницаемости. Наблюдается зависимость вида $1/R$ ширины плазмонного резонанса наночастицы при значениях $R \leq 6\,нм$, но практически без изменения частоты резонанса. Для пленки толщиной $6\,нм$, по-видимому, следует брать $Im\varepsilon_2$ увеличенным в $1.5 - 2$ раза в сравнении со значением для массивных образцов, для пленки $2\,нм$ рост $Im\varepsilon_2$, по-видимому, еще больше. Поэтому в расчете, результат которого приведен на рис.4 *d*, значение $Im\varepsilon_2$ выбрано увеличенным в $5$ раз против значения $0.197$.

Введем параметр $d_{\min}$ размерности длины, характеризующий достижимое разрешение в плоскости изображений. Значение $d_{\min}$ определяется максимальным $k_x = k_x^{\max}$ волн, участвующих в построении изображения. При описанной выше методике построения изображения $k_x^{\max} = k_a$. При $\varepsilon = -1$ оказывается, что $k_x^{\max} = \infty$, а разрешение бесконечно большим. Однако при $2\pi / k_x^{\max}$, приближающемся по значению к расстоянию между атомами в решетке металла и к среднему расстоянию между электронами проводимости в металле, металл более нельзя рассматривать как однородную среду. Перестает работать модель диэлектрической проницаемости, которой мы пользовались до сих пор. Вопрос о том, что представляет собой металл для электромагнитного поля оптической частоты и с пространственной периодичностью, более высокой, чем периодичность решетки металла, является предметом специального исследования. Его мы представим в другом месте.

Сконструируем устройство, в котором будет осуществляться проецирование пространственно-ограниченного поля из одной плоскости в другую плоскость. Это устройство может состоять из двух параллельных полосковых диэлектрических волноводов. Щели на обращенных друг к другу границах этих волноводов будут осуществлять связь между волноводами. Введение слоя из серебра будет усиливать связь первого волновода со вторым и осуществлять пространственно определенное возбуждение одного волновода излучением из другого волновода. Такую же связь можно осуществить и между двумя металлическими полосковыми волноводами, поддерживающими распространение поверхностных плазмонов. В сложном устройстве, состоящем из множества слоев, проецирование поля с помощью серебряного слоя позволит осуществить разветвленные и селективные связи между слоями.

Подведем итог. На пленке металла с отрицательным $\varepsilon$, находящейся в вакууме, существуют поверхностные волны - плазмоны. Рассмотрены поперечно-магнитные волны. Имеются два значения волнового числа плазмона, $k_{sx}$ и $k_{ax}$, соответствующие двум резонансным волнам – симметричной и антисимметричной. Волновое число $k_{sx}$ находится вблизи значения $k_0 = \omega/c$, а $k_{ax}$ сильно превышает это значение. Отличие $k_{ax}$ от $k_{sx}$ тем больше, чем ближе значение $\varepsilon$ к $-1$. По абсолютной величине $\varepsilon$ металла может быть произвольным, как меньше $1$, так и больше $1$. При $-1 < \varepsilon < 0$ плазмон с высокой пространственной частотой также является симметричным.

Пленка металла подобно линзе осуществляет проецирование объекта в плоскость изображения. В отличие от традиционной линзы изображение за металлической пленкой имеет повышенное разрешение, так как содержит детали размером много меньше длины. Эффект наблюдается для металла с $\varepsilon = -1$, $\varepsilon < -1$ и $-1 < \varepsilon < 0$. Наиболее высокое разрешение возникает при $\varepsilon = -1$. Обсужден вопрос о построении изображения с деталями размером менее расстояния между атомами в решетке металла.

Диэлектрическая проницаемость $\varepsilon$ для плазмонов является величиной относительной, равной отношению $\varepsilon_{metal} / \varepsilon_{dielectric}$. Металл с $|\varepsilon| \gg 1$, помещенный в диэлектрик с близкой по значению положительной диэлектрической проницаемостью, будет вести себя как металл с $\varepsilon \approx -1$ в вакууме. Это позволяет осуществить случай $\varepsilon \approx -1$ на произвольной длине волны. Все сказанное о пленке металла в диэлектрике в равной мере относится к пленке диэлектрика в металле.

*Об импульсе фотона в поверхностном плазмоне*

При рассеянии рентгеновских и $\gamma$ - квантов на электроне происходит заметное изменение частоты рассеянного кванта. В процессе рассеяния электрон получает импульс отдачи, а рассеянный квант теряет импульс и, соответственно, энергию. На оптических частотах эффект мал в силу малости импульса оптического кванта.

Выше мы имели в виду квант (фотон) в плоской волне свободного пространства. Фотон в волне в диэлектрической среде имеет импульс, увеличенный в $n$ раз, $n$ - показатель преломления среды. Ниже будет показано, что фотон в поверхностном плазмоне имеет импульс, увеличенный в число раз, равное замедлению плазмона по отношению к волне в свободном пространстве. Замедление может достигать тысячи крат, соответственно импульс фотона в плазмоне в тысячу раз больше.

В свободном пространстве, в диэлектрической среде и в плазмоне импульс фотона равен $\hbar\vec{k}$, $\vec{k}$ - волновой вектор соответствующей плоской волны, однородной или неоднородной. Это утверждение на первый взгляд представляется совершенно очевидным. Однако как оказалось, вопрос об импульсе электромагнитной волны в среде обсуждается уже 100 лет. Эта дискуссия начинается со статей Минковского 1908 г. /19/ и Абрагама 1910 г. /20/. Эти статьи содержат различающиеся выражения для тензора энергии-импульса электромагнитной волны в среде. На протяжении лет были выдвинуты многочисленные аргументы в пользу каждой из сторон. Было также признано, что эксперимент не может дать возможность сделать выбор между двумя точками зрения. Среди авторов, писавших на данную тему, отметим Дж. П. Гордона /21/, Д. В. Скобельцына /22/ и Р. Пайерлса /23/. Представление о современном состоянии дискуссии можно получить с помощью работы /24/. Итог заключается в следующем. Тензор энергии-импульса для электромагнитной волны сам по себе не имеет необходимой полноты. Необходим дополнительный учет тензора энергии-импульса материальной среды. При этом предсказания всех предложенных теорий оказываются одинаковыми, а предпочтительная форма теории оказывается делом персонального выбора автора.

В том, что следует ниже, будем следовать представлениям, развитым в работе Нельсона /25/ и Нельсона и др. /26/. Содержание этих работ заключается в выводе законов сохранения для энергии и импульса электромагнитной волны в диэлектрической среде на основе общих принципов. В качестве последних используются свойства однородности и изотропии времени и пространства, из которых следуют физические законы сохранения. Однородность и изотропия означают, что лагранжиан системы не меняется при изменении начала отсчета времени и при переносе и повороте системы координат в пространстве. Возникающие законы сохранения дают выражения для энергии, импульса и др.

Нельсон показывает, что существуют две импульсоподобные величины – импульс поля, взаимодействующего с веществом, и импульс вещества, подвергающегося действию поля. Закон сохранения импульса поля возникает в силу однородности пространства, импульса вещества - в силу однородности вещества. Импульс вещества назван псевдоимпульсом. Сумма импульса поля и псевдоимпульса дает импульс, названный импульсом волны в диэлектрике. В пересчете на один фотон импульс волны оказывается равным $\hbar\vec{k}$, $\vec{k}$ - волновой вектор волны в веществе.

Результаты работ /25,26/ непосредственно переносятся в задачу об импульсе фотона в поверхностном плазмоне. Это оказывается возможным при следующих изменениях. Поверхностные плазмоны существуют в пространстве с границей раздела двух сред – диэлектрика и металла. Будучи неоднородным 3-хмерным пространством это пространство как двумерное является однородным. К этому пространству следует применить те процедуры, которые применены в /25,26/ по отношению к 3-хмерному пространству, с заменой 3-хмерных плотностей энергии, импульса и их плотностей потоков на двумерные плотности и плотности потоков. В результате возникают совпадающие результаты, в том числе – импульс фотона $\hbar\vec{k}$.

Сильно замедленные плазмоны существуют в различных структурах. Здесь в качестве примера возьмем тонкий слой металла, $\varepsilon_m < 0$, в диэлектрической среде, $\varepsilon_d > 0$, $\mu_m = \mu_d = 1$. Если $|\varepsilon_m|$, модуль $\varepsilon_m$, почти равен $\varepsilon_d$, но превышает его, а пленка имеет толщину $1-3$ нм, то у металлической пленки появляется сильно замедленный плазмон, антисимметричный по структуре поля, с волновым числом $k$, которое вплоть до тысячи раз больше, чем $\omega/c$. Помимо этого медленного плазмона у пленки имеется и другой плазмон, симметричный плазмон. В данном тексте он не интересен, так его волновое число мало отличается от $\omega/c$ (точнее, от $(\omega/c)\sqrt{\varepsilon_d\mu_d}$). Кроме тонких пленок сильно замедленные плазмоны существуют на тонких металлических нитях (1-мерное однородное пространство) и на металлических сферах и металлических частицах иной формы. В последнем случае плазмон представляет собой стоячую волну.

Напомним, что речь идет об оптических частотах. При $k = 10^3 \omega/c$ длина волны плазмона $\lambda = 10^{-3} \cdot 2\pi c/\omega$ оказывается равной $0.5$ нм для зеленого излучения. Это значение уже не очень сильно отличается от размера элементарной ячейки металла. Вопрос о коротковолновом плазмоне с длиной волны, равной или даже меньше пространственного периода решетки металлического кристалла требует дополнительного изучения.

Теперь мы готовы рассмотреть задачу о рассеянии фотона из плазмона на свободном электроне. Ниже будет рассмотрено пространство с плоской границей раздела металл-диэлектрик. Будут также рассмотрены пространства с двумя и даже с четырьмя границами раздела металл-диэлектрик. Наличие дополнительных плоско-параллельных границ раздела не нарушает однородности двумерного пространства. В этих двумерных пространствах нет зависимости их свойств от продольных координат, и потому они оказываются однородными.

Формула Комптона для измененной длины волны фотона имеет вид

$$\lambda' - \lambda = (2\pi\hbar/mc)(1 - \cos\theta). \qquad (20)$$

Здесь $\lambda$ - длина волны падающего фотона, $\lambda'$ - длина волны рассеянного фотона, $m$ - масса электрона, $c$ - скорость света, $\theta$ - угол между волновыми векторами падающего и рассеянного фотонов; $2\pi\hbar/mc = 2.4263 \cdot 10^{-10}$ см для электрона. При рассеянии оптического фотона длины волны $500$ нм сдвиг составляет около $5 \cdot 10^{-4}\%$.

Фотон в поверхностном плазмоне имеет импульс, увеличенный в число раз, равное $n$ - кратности замедления. Поле поверхностной волны (плазмона) находится как внутри металла, так и вне его, оба – вблизи поверхности раздела. При удалении от поверхности раздела поля убывают экспоненциально. При появлении электрона вблизи поверхности металла может произойти рассеяние фотона из плазмона на электроне. Рассеяние может произойти в любую моду рассматриваемого пространства, в том числе – в плоскую волну в прилегающем к металлу диэлектрике. Нас будет интересовать рассеяние в поверхностный плазмон. При рассеянии в плазмон происходит заметное изменение импульса и энергии рассеянного фотона. Расчет показывает, что имеют место следующие приближенные соотношения между энергиями падающего и рассеянного фотонов и для импульса, приобретаемого электроном, который первоначально покоился. Формулы получены при условии, что $n^2(\hbar\omega/2mc^2) << 1$. Ниже штрихом отмечены величины, относящиеся к фотону и электрону после рассеяния. Импульсы обоих фотонов лежат в одной плоскости, кратности замедления для падающего и рассеянного фотонов, $n$ и $n'$, приняты одинаковыми.

$$\hbar\omega' \approx \hbar\omega[1 - n^2(\hbar\omega/2mc^2)(1 - \cos\theta)], \quad p_{e'} \approx n(\hbar\omega/c)(1 - \cos\theta). \qquad (21\ 16)$$

Наибольший импульс электрон получает при $\theta = \pi$, то есть при рассеянии назад. При $\hbar\omega' \approx \hbar\omega$ имеем $p_{e'} \approx 2n\hbar\omega/c$.

Приведенные формулы возникают в результате следующего рассмотрения. При рассеянии имеют место законы сохранения энергии и импульса:

$$E_\gamma + E_e = E_{\gamma'} + E_{e'}, \qquad (22\ 17)$$

$$\vec{p}_\gamma + \vec{p}_e = \vec{p}_{\gamma'} + \vec{p}_{e'}. \qquad (23\ 18)$$

Удобно считать, что первоначально электрон покоился, то есть $E_e = mc^2$, $\vec{p}_e = 0$. Энергия электрона после рассеяния равна $E_{e'} = \sqrt{(\vec{p}_{e'}c)^2 + (mc^2)^2}$. Из (22) следует

$$(\vec{p}_{e'})^2 = [(\hbar\omega - \hbar\omega')^2 + 2(\hbar\omega - \hbar\omega')(mc^2)]/c^2, \qquad (24\ 19)$$

а из (23) –

$$(\vec{p}_{e'})^2 = [n^2(\hbar\omega)^2 + n'^2(\hbar\omega')^2 - 2nn'\hbar\omega\hbar\omega'\cos\theta]/c^2. \qquad (25\ 20)$$

Приравнивая (24) и (25) после алгебраических преобразований получаем уравнение относительно $\hbar\omega'$

$$(\hbar\omega')^2 + 2\hbar\omega'(mc^2/n'^2)[1 - nn'(\hbar\omega/mc^2)\cos\theta] + (mc^2/n'^2)[n^2(\hbar\omega)^2/mc^2 - 2\hbar\omega] = 0. \qquad (26\ 21)$$

Заметим, что при $n, n' \approx 10^3$, $\hbar\omega = 2\,эВ$, $mc^2 = 0.51 \cdot 10^6\,эВ$ (энергия покоя электрона) комбинация $nn'(\hbar\omega/mc^2)$ равна $4$, то есть не мала в сравнении с $1$. При $n, n' = 1$ из (26) следует формула Комптона. Решение уравнения (26) имеет вид

$$\hbar\omega' = -mc^2(n'^2-1)^{-1}[(1-nn'\cos\theta)(\hbar\omega/mc^2)+1] \pm mc^2(n'^2-1)^{-1} \times$$
$$\times \sqrt{1+2(\hbar\omega/mc^2)[(1-nn'\cos\theta)+(n'^2-1)]+(\hbar\omega/mc^2)^2[(1-nn'\cos\theta)^2-(n^2-1)^2]}.$$
(27 22)

Физически приемлемое решение возникает при выборе знака плюс перед корнем в (27). На рис.11 показано это решение для $n = n' = 300$, $\hbar\omega = 2\,эВ$ в полярных (слева) и прямоугольных (справа) координатах. Можно видеть, что при выбранных значениях параметров в (27) энергия рассеянного назад фотона вдвое меньше энергии падающего фотона.

Представляет практический интерес процесс с рассеянием фотона из плоской волны свободного пространства в поверхностный плазмон. Устройство, в котором реализуется такой процесс, может служить как устройство для возбуждения поверхностных плазмонов или как устройство регистрации электронов.

Уравнение для $\hbar\omega'$ получим из (26), положив в нем $n = 1$. Физически приемлемо решение со знаком плюс перед корнем. На рис.12 показана зависимость $\hbar\omega'(\theta)/\hbar\omega$ для $n = 300$, на рис.13 – диаграмма импульсов.

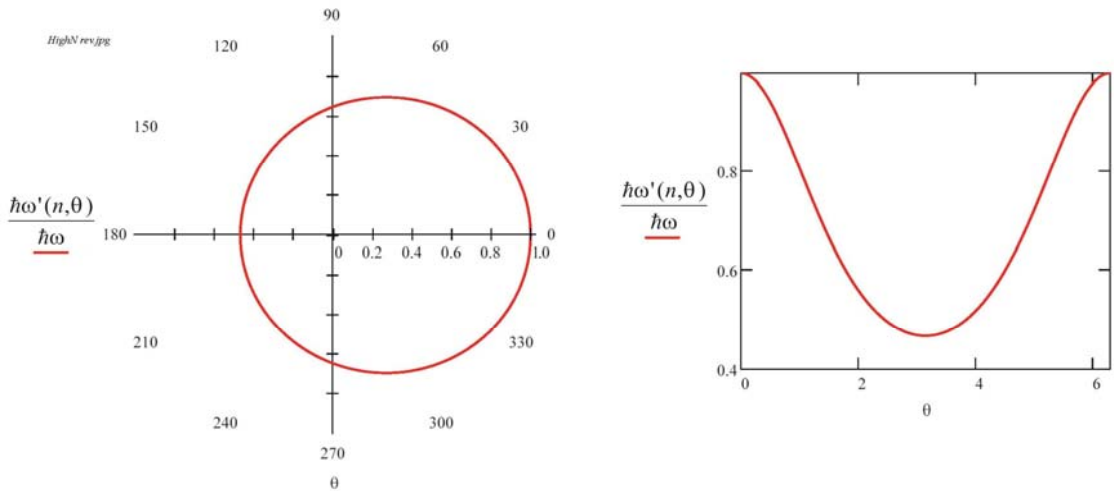

Рис.11. Решение уравнения (22) для $n = n' = 300$, $\hbar\omega = 2\,эВ$.

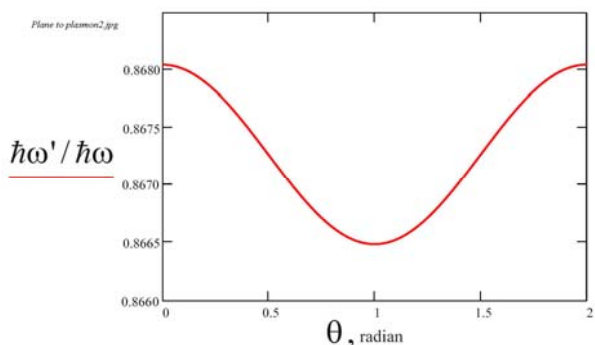

Рис.12. График функции $\hbar\omega'/\hbar\omega$ со знаком плюс перед корнем.
$\hbar\omega = 2\,eV$, $mc^2 = 0.51 \cdot 10^6\,eV$, $n = 300$.

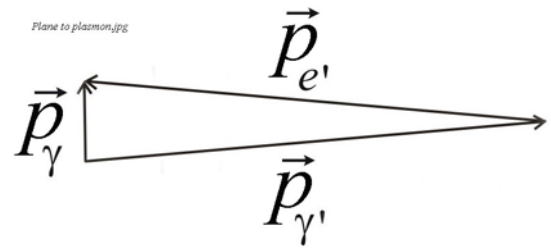

Рис.13. Диаграмма импульсов при рассеянии на электроне фотона из свободного пространства в поверхностный плазмон, $\theta \approx \pi/2$.

Для выбранных параметров частота плазмона приблизительно на $13\%$ меньше частоты падающей волны. Если бы рассеяние происходило на более легкой, чем свободный электрон частице, например, на электроне в зоне проводимости полупроводника, чья эффективная масса $m_{eff}$ меньше массы свободного электрона, то изменение частоты рассеянного фотона было бы более заметным. При $m_{eff} = 0.2m$ отношение $\hbar\omega'/\hbar\omega$ было бы равно приблизительно $0.64$.

Рассмотренный процесс напоминает ядерный распад. Система, состоящая из покоящегося электрона и фотона из плоской волны свободного пространства, имеющая очень малый импульс, при распаде порождает фотон и электрон, оба с большими противоположно направленными импульсами.

Получаемая из (27) пара значений $\hbar\omega'$ и $p_{\gamma'}$ в общем случае не соответствуют резонансному плазмону на частоте $\omega'$. Нерезонансные плазмоны могут возбуждаться на границах раздела металл-диэлектрик, но при этом среди волн в пространстве с диэлектриком у границы должна присутствовать прилегающая волна, нарастающая при удалении от границы раздела. На первый взгляд это обстоятельство запрещает возможность рассеяния в нерезонансный плазмон. Однако это не так. Следует предположить, что на удалении от пленки металла находится вторая, параллельная пленка металла. Расстояние между пленками так велико, что можно не учитывать их взаимное влияние при расчете собственных полей мод. Однако с точки зрения возможного возбуждения плазмона во второй пленке наличие связи существенно. И хотя эта связь мала, ее малость проявляется лишь в малости величины сечения рассеяния фотона, но не препятствует возбуждению плазмона во 2-й пленке. Наличие дополнительных плоско-параллельных границ раздела не нарушает однородности двумерного пространства. В таком двумерном пространстве нет зависимости его свойств от продольных координат, и потому оно по-прежнему является однородным.

Подведем итог. Фотон в поверхностном плазмоне имеет импульс, увеличенный в число раз, равное замедлению плазмона по отношению к фотону той же частоты в плоской волне в свободном пространстве. При рассеянии на свободном электроне поверхностной волны (плазмона) в поверхностную же волну (плазмон) электрон получает импульс, заметно увеличенный по сравнению с импульсом, получаемым при рассеянии плоской волны в плоскую волну, обе - в свободном пространстве. Эффект сходен с эффектом Комптона, обычно наблюдаемым при рассеянии рентгеновских и $\gamma-$квантов. Интерес представляет процесс рассеяния плоской волны в плазмон, при котором электрон получает импульс, многократно превышающий импульс фотона плоской волны свободного пространства. Импульс этого электрона приблизительно равен импульсу фотона в плазмоне, а последний имеет энергию почти равную энергии падающего фотона и импульс, увеличенный в число раз, равное замедлению. Рассмотренные явления могут иметь практическое применение в устройствах нанооптики, например, в устройстве регистрации электронов по плазмонам, возбуждаемым в прилегающих металлических нанопленках и нанонитях, либо для регистрации фотонов.

В заключение можно сказать, что очень медленные поверхностные плазмоны с одной стороны являются очень интересным физическим объектом, а с другой стороны являются ключевым объектом устройств нанооптики, претерпевающей в данный момент интенсивное развитие.